**Manipulations of individual molecules by scanning probe microscopy**

O. Dudko[a], A.E. Filippov[b], J. Klafter[a] and M. Urbakh[a]

[a]School of Chemistry, Tel Aviv University, 69978 Tel Aviv, Israel.

[b]Donetsk Institute for Physics and Engineering of NASU, 83144, Donetsk, Ukraine

**Abstract**

In this Letter we suggest a new method of manipulating individual molecules with scanning probes using a "pick-up-and put-down" mode. We demonstrate that the number of molecules picked up by the tip and deposited in a different location can be controlled by adjusting the pulling velocity of the tip and the distance of closest approach of the tip to the surface.

**Corresponding author:** Prof. Michael Urbakh, Address: School of Chemistry, Tel Aviv University, 69978,Tel Aviv, Israel, E-Mail: urbakh@post.tau.ac.il

Phone: +972-3-6408324 ; Fax: +972-3-6409293

*Introduction.*

Soon after the invention of the scanning tunneling microscopy (STM) and the atomic force microscopy (AFM) it has been recognized that scanning can alter surface topography. That time this was considered as a drawback for imaging. However, these observations led to an idea of controllable modification of surface structure at the atomic scale which attracted the attention of a large number of research groups [1-4]. The ability to manipulate individual atoms, molecules and clusters with scanning probes has opened new fascinating areas of research and allowed to perform "engineering" operations at the ultimate limits of fabrication.

Manipulations are usually classified into two types: lateral and vertical [2-4]. In the lateral case an object is displaced (pulled, pushed or slides) from one position to another on the surface, while in the vertical case the object is transferred between the surface and the tip. The vertical mode is also referred as "pick-up-and-put-down" [3]. Lateral movements of adsorbates have been a subject of numerous experimental and theoretical studies [5-9]. However, controllable vertical manipulations of individual adsorbates by AFM and STM are just at their early stages [3,4,10-12]. It is more difficult to control vertical manipulations than lateral ones, since the energy barriers needed to be overcome when pulling an individual adsorbate off a surface are usually higher than for lateral movements.

When tip is brought into the close vicinity of a surface, the two potential wells, corresponding to the equilibrium position of the adsorbate on the tip or on the surface when they are far apart, overlap (see Fig.1). As a result, the barrier for a transfer of the adsorbate between the surface and the tip decreases. The remaining barrier can be crossed spontaneously due to the presence thermal fluctuations. However, since the relaxation times compete with the moving tip, adsorbate cannot always follow spontaneously the motion of the tip, and a probability of the adsorbate transfer between the surface and the tip depends not only on the tip proximity but also on its velocity. In this Letter we demonstrate that the number of molecules picked up by the tip and deposited on another surface can be controlled by adjusting the pulling velocity of the tip and the distance of closest approach of the tip to the surface. This differs from an earlier suggestion to control the extraction of atoms from a surface through the duration of maximal approach and tip displacement towards the surface [10].

*The model.*

In order to mimic the manipulation of adsorbates by scanning probe microscopy we introduce a model which consists of a monolayer of $N$ interacting molecules with masses $m$ and coordinates $\mathbf{r}_i = \{x_i, y_i, z_i\}$ located on a substrate, and a tip of mass $M$ and center-of mass coordinate $\mathbf{R}_i = \{X_i, Y_i, Z_i\}$. The tip is pulled by a spring of stiffness $K$ in the $z$-direction perpendicular to the surface $\{x, y\}$. The spring is connected to a stage which moves with a constant velocity $\mathbf{V}$. The dynamics of this system is described by a system of $3N+3$ equations of motion for the tip and the molecules:

$$M \partial^2 \mathbf{R}/\partial t^2 + G\, \partial \mathbf{R}/\partial t + \sum_{i=1}^{N} \partial U^{t\text{-}m}(\mathbf{r}_i\text{-}\mathbf{R})/\partial \mathbf{R} + \partial U^{t\text{-}s}(\mathbf{R})/\partial \mathbf{R} + K(\mathbf{R}\text{-}\mathbf{V}t) = 0 \qquad (1)$$

$$m\, \partial^2 \mathbf{r}_i/\partial t^2 + g\partial \mathbf{r}_i/\partial t + \partial[U^{t\text{-}m}(\mathbf{r}_i\text{-}\mathbf{R}) + U^{m\text{-}s}(\mathbf{r}_i)]/\partial \mathbf{r}_i + \sum_{i \neq j}^{N} \partial U^{m\text{-}m}(\mathbf{r}_i\text{-}\mathbf{r}_j)/\partial \mathbf{r}_j = f_i(t),$$

$$i=1,\ldots,N. \qquad (2)$$

Here the potentials $U^{m-m}$, $U^{m-s}$, $U^{t-m}$ and $U^{t-s}$ describe molecule-molecule, molecule-substrate, molecule-tip and tip-substrate interactions respectively. The parameters $G$ and $g$ account for the dissipation of the kinetic energy of the tip and each molecule, correspondently. The effect of the thermal motion of the adsorbates is given in terms of a random force $f_i(t)$, which is $d$-correlated, $<f_i(t)f_i(0)>=2mk_BgTd(t)d_{ij}$. $T$ is the temperature, $k_B$ is the Boltzmann constant.

In our numerical simulations the molecule-molecule and tip-molecule interactions have been modeled by Morse potentials

$$U^{m-m}(r_i - r_j) = U_0^{m-m}\left\{\left[1 - \exp(-2b^{m-m}(r_i - r_j - R_C^{m-m})/R_C^{m-m})\right]^2 - 1\right\}, \quad (3)$$

$$U^{t-m}(R - r_i) = U_0^{t-m}\left\{\left[1 - \exp(-2b^{t-m}(R - r_i - R_C^{t-m})/R_C^{t-m})\right]^2 - 1\right\}, \quad (4)$$

while for $U^{m-s}$ and $U^{t-s}$ we used

$$U^{m-s} = \left\{U_0^{bulk} + U_0^{surf}\left[\cos(ax) + \cos(ay)\right]\right\}\exp[-(z - l_z)^2/d^2], \quad (5)$$

$$U^{t-s} = C_0 \exp[-(Z - L_z)^2/c_0^2], \quad (6)$$

where $U_0^{t-m}$, $b^{t-m}$, $R_C^{t-m}$, $U_0^{m-m}$, $b^{m-m}$, $R_C^{m-m}$ are the parameters of the Morse potential. It was also taken into account that the dissipation $g$ decreases when the molecules move away from the surface, $g(z) = g_0[1 + \exp(-z^2/d^2)]$. It should be emphasized that our further conclusions are mostly independent of the particular forms of the potentials $U^{t-m}$, $U^{m-s}$ and $U^{t-s}$.

*Qualitative consideration of nano-manipulation.*

Qualitative features of the suggested mechanism of manipulation of individual molecules can be understood within the framework of a simplified one-dimensional model. The model describes a single particle located on the uniform surface and interacting with the tip, which is pulled out off the surface with a constant velocity $V = \dot{Z} = const$, starting from a height $Z_0$. Equations (1)-(2) reduce to a 1D equation of motion for the position of the molecule, $z$, which under overdamped conditions, $g \gg 1$, reads

$$dz/dZ = -\left(\partial U_{eff}/\partial z\right)/gV. \qquad (7)$$

Here $U_{eff} = U^{m-t}(z,Z) + U^{m-s}(z)$ is an effective potential experienced by the molecule due to the surface and tip. The coordinate $Z$ of the tip enters as a parameter.

Typical evolution of the potential $U_{eff}$ with an increase of tip-surface distance $Z$ is shown schematically in Fig.1. When the tip and surface are in close contact the two wells corresponding to adsorption on the tip or on the surface overlap, and the resulting $U_{eff}$ can attain a form of an asymmetric single-well. With an increase in the tip-surface distance the effective potential $U_{eff}$ takes the form of two-well potential, and the barrier between two minima grows.

The set of solutions of Eq. (7) for different values of $Z_0$ and $V$ presents trajectories in the coordinates $(z, Z)$, which give a phase portrait of the dynamical system in the space of parameters of the energy functional $U_{eff}(z,Z)$. Typical phase portraits are shown in Fig.2 for four values of pulling velocities $V$. All trajectories in Fig.2 can be separated into two types: (1) trajectories that correspond to a regime where the molecule remains on the surface, and 2) trajectories that belong to a regime in which the tip picks up the molecule and drives it away from the surface. For the first type of trajectories $z \to z_{ad}^s$ when $Z \to \infty$, and for the second type of the trajectories $z \to Z - z_{ad}^t$ when $Z \to \infty$, where $z_{ad}^s$ and $z_{ad}^t$ are the molecule-surface and molecule-tip distances for the cases of equilibrium adsorption at the substrate (in the absence of tip) and at the tip (in the absence of substrate). Fig.2 shows that even being trapped by the tip at small $Z$, the molecule cannot always follow the tip motion. Due to a finite relaxation time $1/g$ this depends on the tip velocity $V$. For high pulling velocities the molecule always remains at the surface, independent of the starting tip position, $Z_0$ (see Figs.2a). As $V$ decreases, the second type of solutions sets in (Fig.2 b-d). Furthermore, the starting position of the tip, $Z_0$, for which the molecule can still be picked up by the tip increases with a decrease of the pulling velocity.

The above consideration allows to define the $Z_0$-dependent critical velocity of the tip, $V_{cr}(Z_0)$, that is the maximal $V$ for which the tip drives the particle away from the surface. The result is presented in Fig.3. For all values of $V$ and $Z_0$ lying below the curve $V_{cr}(Z_0)$ the tip does pick up the molecule, and for the values above the curve the molecule does not follow the tip and remains on the surface.

The largest allowed value of the critical velocity can be estimated analytically. In order to do this we consider a motion of the molecule which is trapped by the tip, and assume that the distance between the molecule and the tip remains constant, $z_* = Vt - z(t) = const$, when the tip is driven away from the surface. In this regime an effective potential experienced by the molecule is dominated by the attraction to the tip and it can be approximated by $U_{eff} \approx \tilde{e} \exp[-(z-Vt)^2/\tilde{s}^2]$. Under these conditions the equation of motion (7) leads to the following relation between $z_*$ and V:

$$1 - (\tilde{e}/gV) z_* \exp(-z_*^2/\tilde{s}^2) = 0 \qquad (8)$$

Equation (8) has a solution only for $z_* < \tilde{s}/\sqrt{2}$ and $V < V_{cr}^* = \tilde{e}\tilde{s}/(g\sqrt{2e})$. For $V > V_{cr}^*$ the molecule cannot follow the tip motion and remains at the surface. Thus $V = V_{cr}^*$ is the maximal driving velocity for which the tip can pick up the molecule. The estimated value of $V_{cr}^*$ is in good agreement with the numerical results presented in Fig.3.

The dependence $V_{cr}(Z_0)$ does not only give a clue of how to manipulate single molecules but also allows to estimate a range of tip velocities, for which the tip picks up a desirable number, $N_{tr}$, of molecules when it is driven away starting at the distance $Z=Z_0$. In order to do this we define a function $Z_{cr}(V)$ which gives the maximal value of the initial tip-surface distance for which the tip can trap the molecule being driven away with a velocity $V$. The function $Z_{cr}(V)$ is the reciprocal of the function $V_{cr}(Z_0)$. Using this information we can conclude that the tip will pick up all molecules located under the tip within a circle of the radius $R = \sqrt{Z_{cr}^2(V) - Z_0^2}$ (Fig.4). Here we assumed that molecules are distributed uniformly on the surface and do not interact among themselves. Taking into account that $R \propto d\sqrt{N_{tr}}$, where $d$ is an average distance between adsorbates on the surface and $N_{tr}$ is a number of molecules located within the circle, we obtain a following relation between a pulling velocity and a number of molecules picked up by the tip

$$Z_{cr}(V) = \sqrt{Z_0^2 + d^2 N_{tr}}. \qquad (9)$$

Thus intersections of the curve $V_{cr}(Z_0)$ with vertical lines $Z=\sqrt{Z_0^2+d^2 N_{tr}}$ for $N_{tr}=1,2,3...$ which are shown in Fig.3 give the maximal tip velocity for which the tip picks up *a given number of particles*, $N_{tr}$, when it is driven away from the surface starting at distance $Z_0$.

It should be noted that the range of tip velocities suitable for controllable molecular manipulation strongly depends on the interaction of the tip with the molecules. The latter can be made adjustable by modifying the tip chemically [13]. In this way the critical velocity can be moved into the desirable range. Surprisingly the characteristic time to extract a molecule has been found to be as slow as 10 ms [10], a time that allows the tip velocity to act as a control parameter.

The same mechanism of manipulation by adjusting the tip velocity and the distance of the closest approach to the surface can be used for a deposition of a given number of molecules on the surface. Below we illustrate the proposed mechanism of pick-up-and-put-down mode of manipulation by of numerical simulations.

*Results of the simulation and discussion.*

We have performed numerical simulations of Eqs. (1)-(2) that describe the coupled dynamics of the externally driven tip and the monolayer of adsorbed molecules. Solving the equations we started from the equilibrium configuration produced when the tip is brought into close contact to the surface. Then the tip was pulled away from the surface by the spring with a constant velocity. The number of molecules picked up by the tip has been found repeatedly. As a result we obtained a map of probability to trap a given number of particles by the tip at a given driving velocity, which is presented in Fig. 5. Regions of high and low probability are displayed by red and blue colors correspondingly. Figure 5b presents the distribution functions of the number of trapped particles for three representative velocities. The map shows that the number of molecules picked up by the AFM tip can vary over a wide range; for the parameters used here this number varies from 0 to 8. The desirable number can be achieved by tuning the driving velocity. In accordance with the qualitative picture discussed above, the number of trapped molecules decreases with the increase in the driving velocity.

It should be noted that not all possible numbers of molecules can be trapped with equal probability. The probability map demonstrates that there are "preferred" numbers of molecules (1, 3, 5, 8) which can be picked up with a high probably, while trapping of 4, 6 and 7 molecules is less probable. The origin of such "magic numbers" can be explained by analyzing molecular configurations which can be formed around the tip. Fig.6 presents examples of the energetically preferred configurations which have been observed in the simulations: five particles (four in a plane and fifth atop the tip, Fig.6a), and eight particles (six form hexagonal structure with the tip and two compensate an asymmetry cased by the difference in size of the particles and the tip, Fig.6b). We remark that the shape of these configurations and the number of particles in them are not universal. They are determined by the radius of the tip and parameters such as molecule-tip and molecule-molecule interactions.

The map shows that the changing of the pulling velocity indeed allows to control the number of molecules transferred from the adsorbed layer to the tip. The proposed manipulation can be optimized and further controlled by adjusting the distance of the closed approach of the tip to the surface and a waiting time before the pulling out of the surface.

Figure captions.

1. Schematic presentation of a typical evolution of the effective potential experienced by the molecule with an increase in the tip-surface distance. Parameter values: $U_0^{bulk}/U_0^{t-m} = 0.1$, $U_0^{surf}/U_0^{t-m} = 0.05$, $d/R_C^{t-m} = 2$, $l_z/R_C^{t-m} = 6$. Lengths and energy are in units of $R_C^{t-m}$ and $U_0^{t-m}$ respectively.

2. Trajectories of atoms as a function of the tip coordinates for four values of the pulling velocity: a) V=10; b) V=9.5; c) V=7.5; d) V=0.7. Parameter values: $U_0^{bulk}/U_0^{t-m} = 1.9$, $d/R_C^{t-m} = 10$. Lengths and velocities are in units of $R_C^{t-m}$ and $g_0 R_C^{t-m}/m$ respectively.

3. Maximal velocity for which the tip still drives a particle away *vs* starting height of the tip. For all values $V$ and $Z_0$ below the curve $V_{cr}(Z_0)$ the tip picks the molecule up, while for the values above the curve the molecule remains on the surface. Intersections of the curve $V_{cr}(Z_0)$ with vertical lines give maximal tip velocity for which the tip picks up a given number of particles, $N_{tr}$=1,2,…,7 when driven from stating height $Z_0$. Parameter values: as in Fig. 2.

4. Schematic explanation to the analytical estimation of the critical velocity $V_{cr}$: the tip picks up all the molecules located under the tip within a circle of the radius $R$.

5. (a) Probability map, giving the probability to trap a given number of particles at a given driving velocity of the tip. The bar to the right of the map sets up a correspondence between colors and the probability $P(N_{tr},V)$.

(b) Histograms for number of trapped particles corresponding to three values of velocity. Parameter values: $N=100$, $M=30m$, $G=30\%$, $a/R_C^{t-m}=6.3$, $l_z/R_C^{t-m}=2$, $d/R_C^{t-m}=1$, $KR_C^{t-m}/U_0^{t-m}=4.7$, $U_0^{m-m}/U_0^{t-m}=0.07$, $b^{m-m}=0.6$, $R_C^{m-m}/R_C^{t-m}=1$, $U_0^{bulk}/U_0^{t-m}=0.7$, $U_0^{surf}/U_0^{t-m}=0.05$, $C_0/U_0^{t-m}=-0.8$, $c_0/R_C^{t-m}=2$, $L_Z/R_C^{t-m}=-2$, $k_B T/U_0^{t-m}=10^{-3}$. Lengths and velocities are in units of $R_C^{t-m}$ and $g_0 R_C^{t-m}/m$ respectively.

6. Examples of preferred configurations formed by the molecules around the tip: a) five particles (four in a plane and fifth atop the tip); b) eight particles (six form hexagonal structure with the tip and two compensate an asymmetry cased by the difference in size of the particles and the tip).

**References**

1. Avouris, Ph. *Acc. Chem. Res.* **1995**, *28*, 95.

2. Gimzewski, J.K.; Joachim, C. *Science* **1999**, *283*, 1683.

3. Nyffenegger, R.M.; Penner, R.M. *Chem. Rev.* **1997,** *97*, 1195.

4. Gauthier, S. *Appl. Surf. Sci.* **2000**, *164*, 84.

5. Eigler, D.M.; Schweizer, E.K. *Nature* **1990**, *344*, 524 ; Crommie, M. F.; Lutz, C. P.; Eigler, D. M. *Science* **1993**, *262*, 218.

6. Meyer, G.; Zophel, S.; Rieder, K. H. *Appl. Phys. Lett.* **1996**, *69*, 3185 ; *Phys. Rev. Lett.* **1996**, *77*, 2113.

7. Jung, T. A.; Schlittler, R. R.; Gimzewski, J. K.; Tang, H.; Joachim, C. *Science* **1996**, *271*, 181.

8. Fishlock, T.W.; Oral, A.; Egdell, R.G.; Pethica, J.B. *Nature* **2000**, *404*, 743.

9. Moresco, F.; Meyer, G.; Rieder, K.H.; Tang, H.; Gourdon, A.; Joachim, C. *Appl. Phys. Lett*. **2001**, *78*, 306.

10. Dujardin, G.; Mayne, A.; Robert, O.; Rose, F.; Joachim, C.; Tang, H. *Phys. Rev. Lett.* **1998**, *80*, 3085.

11. Buldum, A.; Ciraci, S. *Phys. Rev. B* **1996**, *54*, 2175.

12. Ben Ali, M.; Ondarcuhu, T.; Brust, M.; Joachim, C. *Langmuir* **2002**, *18*, 872.

13. Wong, S. S.; Joselevich, E.; Woolley, A. T.; Cheung, C. L.; Lieber, C. M. *Nature* **1998**, *94*, 52.

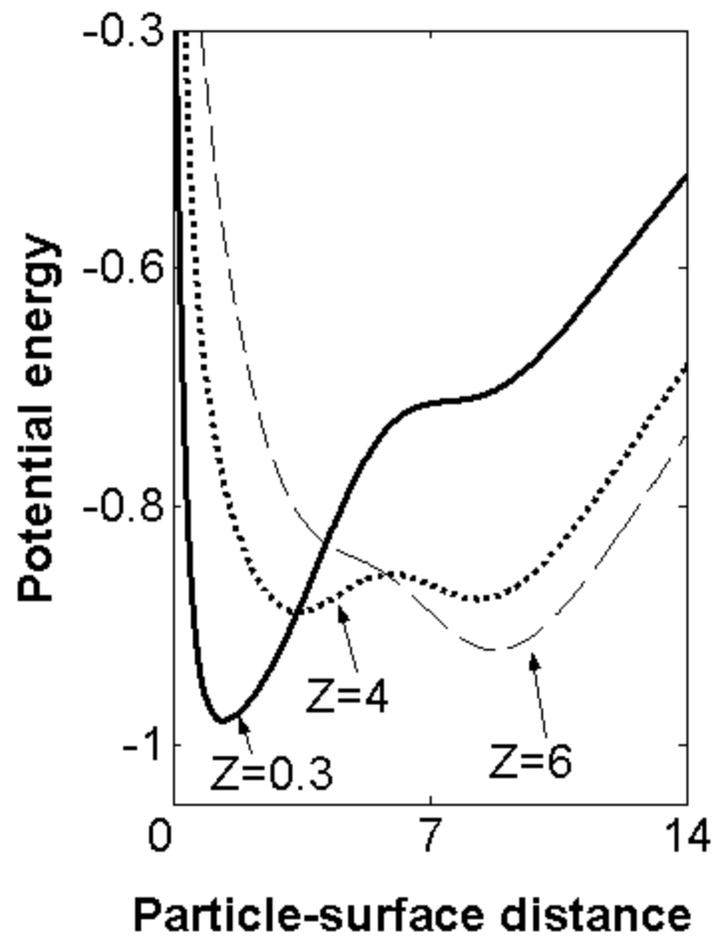

Fig.1

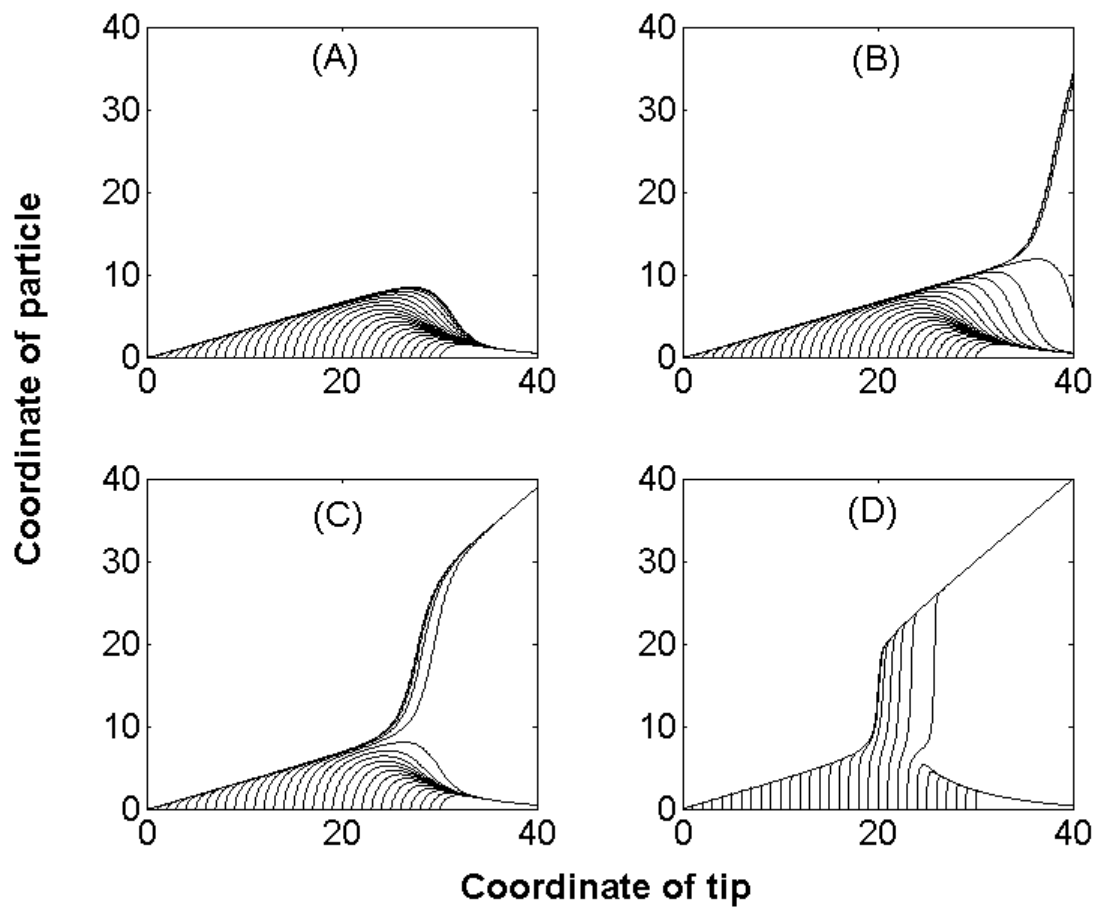

Fig.2

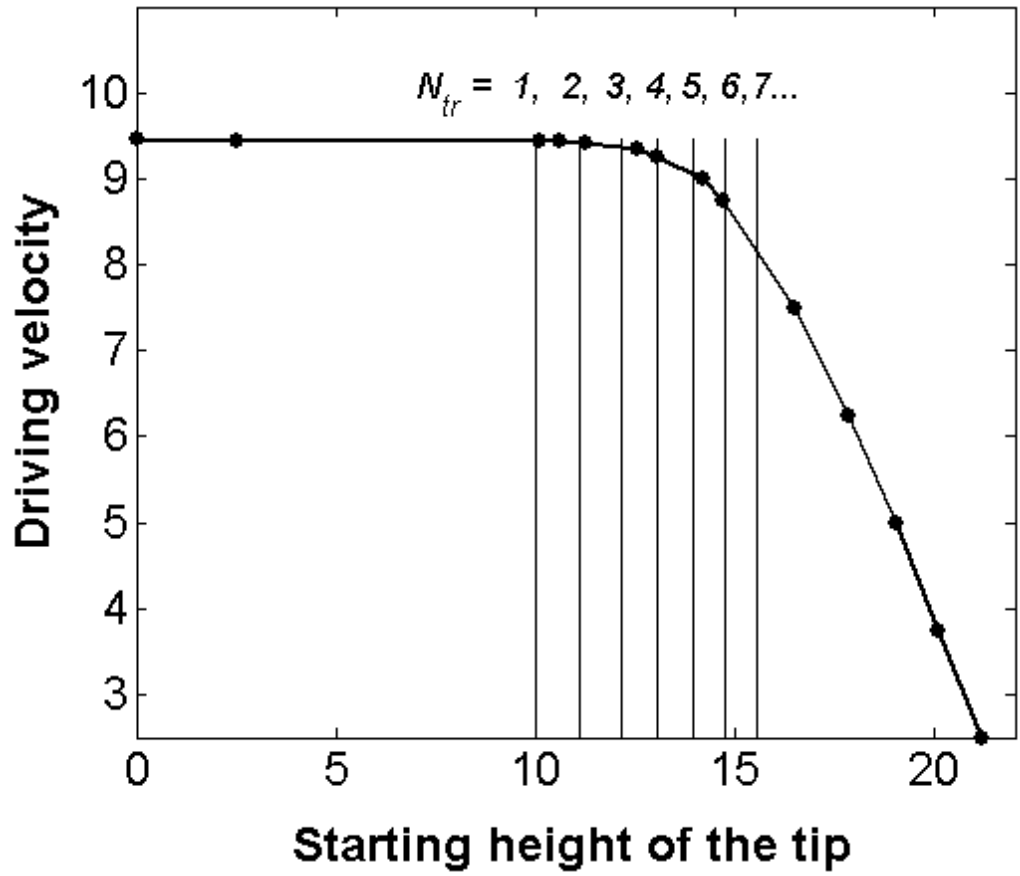

Fig.3

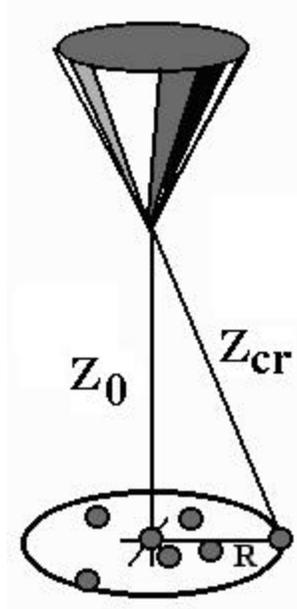

Fig.4

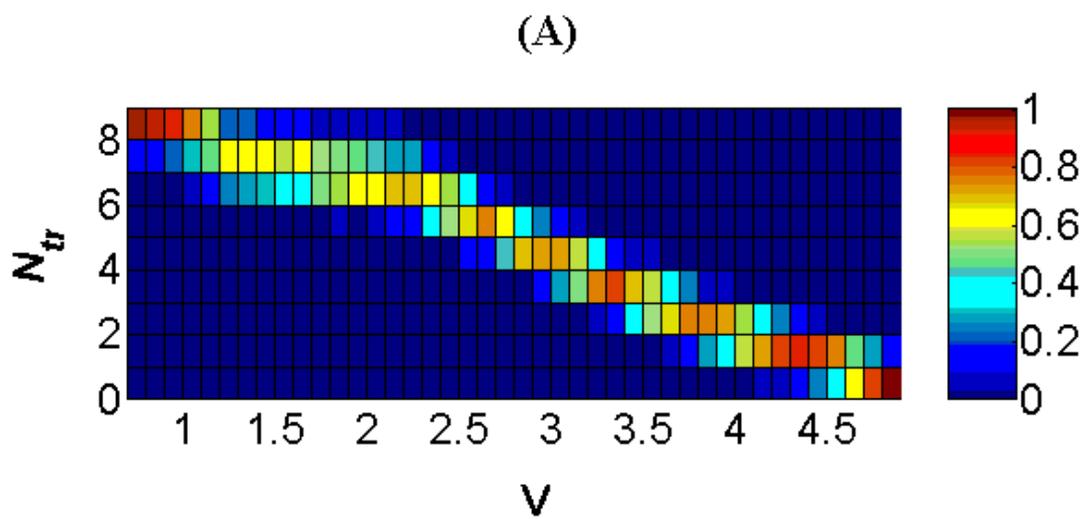

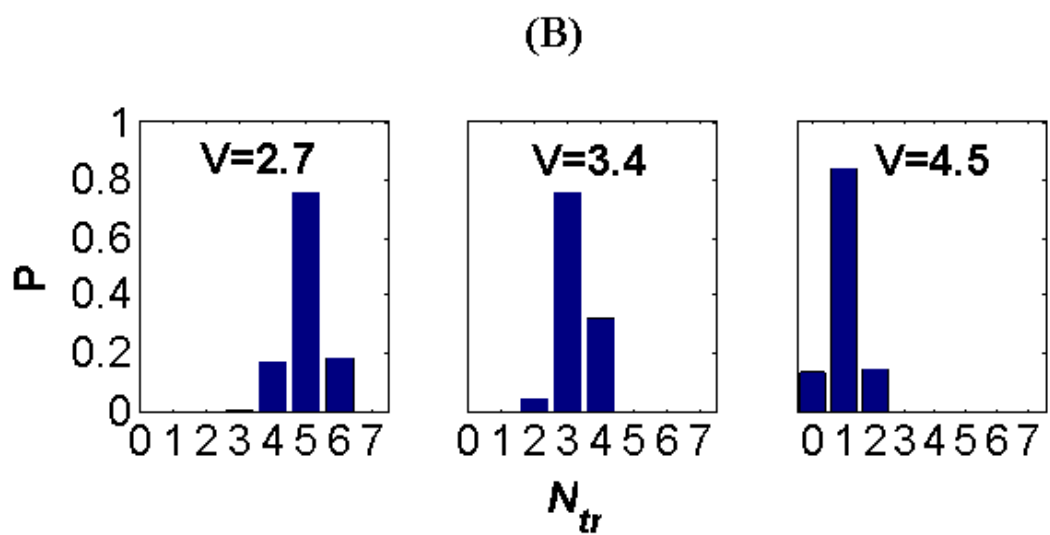

Fig.5

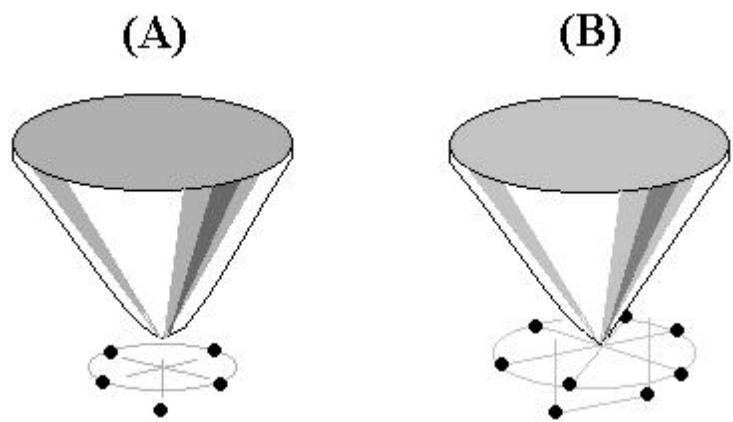

Fig.6